\documentclass[pra, aps, superscriptaddress, twocolumn, showpacs, 10pt, floatfix]{revtex4-1}
\usepackage{amsmath}
\usepackage{amssymb}
\usepackage{graphicx}
\newcommand\sqrtwo{\sqrt{2}}
\newcommand\Mo{\hat{M}}
\newcommand\Eo{\hat{E}}
\newcommand\Uo{\hat{U}}
\newcommand\sio{\hat{\sigma}}
\newcommand\si{\sigma}
\newcommand\Mb{\mathbf{M}}

\newcommand\nv{\vec{n}}
\usepackage{xcolor}

\newcommand{\ket}[1]{\vert#1\rangle}
\newcommand{\bra}[1]{\langle#1\vert}

\begin{document}
\title{Sequential unsharp measurement of photon polarization}    
\author{Peter Adam}
\email{adam.peter@wigner.hu}
\affiliation{Institute for Solid State Physics and Optics, Wigner Research Centre for Physics,\\P.O.\ Box 49, H-1525 Budapest, Hungary}
\affiliation{
Institute of Physics, University of P\'ecs, Ifj\'us\'ag \'utja 6, H-7624 P\'ecs, Hungary}
\author{Lajos Di\'osi}
\affiliation{Institute for Solid State Physics and Optics, Wigner Research Centre for Physics,\\P.O.\ Box 49, H-1525 Budapest, Hungary}
\affiliation{Department of Physics of Complex Systems, E\"otv\"os Lor\'and University,\\1117 Budapest, P\'azm\'any P.\ s.\ 1/A, Hungary}

\begin{abstract}
We propose a general experimental scheme based on binary trees of partially polarizing beam splitters (PPBSs) for realizing sequential unsharp measurements of photon polarization.
The sharpnesses and the bases of the particular photon polarization measurements can be chosen arbitrarily by using corresponding PPBSs and phase plates in the setup.
In the limit of low sharpnesses the scheme can realize sequential weak measurements, too.
We develop a general formalism for describing sequential unsharp measurements of photon polarization in which the particular unsharp measurements are characterized by appropriate measurement operators.
We show that a straightforward experimental realization of this model is the proposed scheme.
In this formalism the output polarization states after the sequential measurement and any correlation functions  characterizing the measurement results can be easily calculated.
Our model can be used for analyzing the consequences of applying postselection and reselection in the measurement.
We derive the anomalous mean value for an unsharp polarization measurement with postselection and the anomalous second-order correlation function for the sequential unsharp measurement of photon polarization with reselection.
We show that these anomalies can be easily measured using the proposed scheme.
\end{abstract}
\maketitle

\section{Introduction}

In modern quantum theory, the notion of measurements has been extended
beyond that of the traditional ideal (sharp) ones, in order to encapsulate coherent unsharpnesses \cite{Krau83,NieChu00}.
An unexpected major motivation for unsharp quantum measurements came with the theoretical discovery of the weak value anomaly.
Such paradoxical measurement outcomes occur in the so-called weak measurements (WMs) meaning asymptotically unsharp measurements, i.e., approaching no-measurement at all \cite{AAV88,Dio06}. 
In WM, the meter of the measuring device is entangled weakly with the system and only a small amount of information is available by detecting the meter's state while the disturbance to the system's initial state $\ket{i}$ remains limited.
The state $\ket{i}$ does not collapse to an eigenstate of the measured observable $\hat A$.
When one postselects a particular quantum state $\ket{f}$ after the WM the statistical mean $\Mb A$ of the measured value $A$ will be the so-called weak value: 
\begin{equation}
\Mb A = A^{WV}=\mathrm{Re}\frac{\bra{f}\hat{A}\ket{i}}{\langle{f}|i\rangle}.
\end{equation}
The weak value anomaly means that $A^{WV}$ may fall outside the spectrum of $\hat A$.
The increased range of possible measurement outcomes inspired the proposal of the idea of using WMs for enhancing the precision in metrology and for amplifying ultra-small physical effects. The advantage of weak-value amplification was demonstrated in several experiments \cite{HostenScience2008, DixonPRL2009, BrunnerPRL2010, XuPRL2013, HallajiNP2017, PiacentiniSciRep2018, XuPRL2020, KrafczykPRL2021}, though the comprehensive theoretical explanation of all the experimental observations is still under exploration \cite{FerriePRL2014, KneePRX2014, DresselRMP2014, ZhangPRL2015, VaidmanPTRSA2017, HarrisPRL2017, RenPRA2020, ArvidssonNC2020, ChenEntropy2021}.

For both theoretical investigations and experimental tests of quantum measurements, photon polarization serves as an ideal system.
Given a laboratory photon along a definite path, a polarization beam splitter (PBS) will maximally entangle the polarization with the path degrees of freedom; the latter can subsequently be measured using a detector.
Thus the path plays the role of meter, taking two well-defined positions after the PBS, indicating horizontal and vertical polarizations, respectively.

The earliest implementation of the measurement of photon polarization weak value required the precise control of the photon's Gaussian wave function in the transverse plane, whose finite width introduced some unsharpness of the path, i.e., of the meter \cite{Ritetal09}.
In this experiment the weak measurement was realized via measuring the polarization-dependent spatial walk-off of the Poynting-vector of the single-photon induced by its propagation through a birefringent medium.
It turned out later that the meter does not have to be continuous: it could be a second photon, weakly entangled with the first one, like in the experiments in Refs. \cite{PrydePRL2005, RozemaPRL2012}.  

Sequential WMs, or, more generally, repeated unsharp polarization measurements on the same photon, however, remained hard to implement experimentally.
Sequential WMs are important for the following reason.
As WMs hardly disturb the quantum system, therefore it is possible to measure non-commuting observables in succession. From the resulting sequential weak values joint properties of the observables can be extracted \cite{Mitetal07,Dio16,Cohen2019}.
Experimental sequential measurements of incompatible, that is, different polarizations of a photon have been reported recently.
The first successful experiment for consecutive weak measurement of two different polarizations was performed by Piacentini et al.~\cite{Piaetal16} based on the method of Ref.~\cite{Ritetal09}. 
Rebufello et al.~\cite{RebufelloLight2021} arranged a sequence of seven unsharp measurements of the same polarization on the same photon, but they detected the sum of the seven outcomes, not the seven outcomes themselves.
For general sequential unsharp polarization measurements, however, discrete paths (and/or ancilla photon's polarizations) can be more promising meters.
Kim et al.~\cite{KimNC2018} realized the measurement of the sequential weak value of two incompatible polarization observables by making use of two-photon quantum interference.
Chen et al.~reported three consecutive WMs of non-commuting polarization observables using the discrete paths of the single photon as meters \cite{Cheetal19}.
The experiment of Foletto et al.~\cite{Foletal21}, with a specific purpose related to WMs, contained two consecutive WMs.
In the latter two experiments, WMs along the path of the photon were performed by interferometric units; each being the same in structure and different by parameters. The function of the units is simple: they realize partially polarizing beam splitters (PPBSs) \cite{Floetal18}.
A PPBS will partially entangle the polarization and the path, allowing for an unsharp polarization measurement when the photon along one of the two paths is detected. 
These experiments took the full statistics of their two/three consecutive unsharp measurements in four/eight separate runs using just two/three interferometric units each time adjusted differently.
Full statistics in one run would require a tree-like structure of interferometric units (just mentioned in \cite{Foletal21}).

In this paper, we develop a general formalism based on appropriate measurement operators for describing sequential unsharp measurements of photon polarization.
This model also describes sequential WMs in the limit of low sharpnesses
and it can be used to calculate any correlation functions characterizing the measurement results.
We show that a straightforward experimental realization of sequential unsharp polarization measurements
is a binary tree structure of PPBSs.
In this scheme, the sharpnesses and the bases of the particular photon polarization measurements can be adjusted arbitrarily.
The scheme is scalable,; it is straightforward to increase the number of consecutive measurements.
It can be also used for realizing unsharp measurements with postselection or reselection. We analyze examples for these cases and derive the anomalous mean value in the former and the anomalous second-order correlation function in the latter case.

The paper is organized as follows.
In Secs.~\ref{UM} and \ref{SUM} the general formalisms for describing an unsharp measurement and the sequential unsharp measurements of photon polarization are presented, respectively.
In Sec.~\ref{SUMreal} we propose the experimental scheme for realizing sequential unsharp measurements of photon polarization.
In Sec.~\ref{UMpsrs} we analyze the consequences of using postselection and reselection in our theory and in the proposed scheme.
Finally, conclusions are drawn in Sec.~\ref{Sec:conc}.

\section{Unsharp measurement of photon polarization}
\label{UM}
The polarization of a single photon can be described in a two-dimensional Hilbert space. The general polarization state of the photon is
\begin{equation}\label{eq:gps:HV}
\ket{\psi}=\alpha\ket{H}+\beta\ket{V},
\end{equation}
where $\ket{H}$ and $\ket{V}$ are the horizontal and vertical polarization states, respectively.
For measuring the polarization one can generally use two other distinguished bases determined by the basis states
\begin{eqnarray}
\ket{L/R} = \frac{1}{\sqrt2} \left( \ket{H} \pm i \ket{V} \right)
\end{eqnarray}
corresponding to the left and right circular polarization, and by the basis states
\begin{eqnarray}
\ket{D/A} = \frac{1}{\sqrt2} \left( \ket{H} \pm \ket{V} \right)
\end{eqnarray}
corresponding to the diagonal and anti-diagonal polarization.
Note that the basis states $\ket{D/A}$, $\ket{L/R}$, and $\ket{H/V}$ are the eigenstates of the Pauli-operators $\sio_x$, $\sio_y$, and $\sio_z$, respectively.
In the case of projective measurement the corresponding projectors can be expressed by the Pauli operators as
\begin{eqnarray}
\label{eq:Pauli1}
\ket{D/A}\bra{D/A}=\frac{1\pm\sio_x}{2},\\
\label{eq:Pauli2}
\ket{L/R}\bra{L/R}=\frac{1\pm\sio_y}{2},\\
\label{eq:Pauli3}
\ket{H/V}\bra{H/V}=\frac{1\pm\sio_z}{2}.
\end{eqnarray}

Next, we consider the unsharp polarization measurement in the H/V basis, that is, the unsharp measurement of the observable $\sio_z$.
As a most general description of this measurement, we introduce the following POVM consisting of two positive operators known as effects $\Eo_\nu$:
\begin{equation}\label{eq:Effects}
\Eo_\pm=\frac{1\pm\cos(2\chi)\sio_z}{2}=\frac{1\pm\sin(2\eta)\sio_z}{2},   
\end{equation}
satisfying $\sum_{\nu}\Eo_\nu=1$, 
where $\chi\in[0,\pi/4]$ is the parameter of unsharpness, while $\eta=\pi/4-\chi$ is the parameter of sharpness.
Note that, for simplicity, we apply here and also in the following the notation $\nu=\pm$ denoting the outcomes $\nu=\pm1$ of the measurement whenever $\nu$ is used as an index, and for convenience, we use the parameters $\eta$ and $\chi$ alternatively.
Projective measurement corresponds to $\chi=0$, weak measurement \cite{AAV88} corresponds to $\eta\rightarrow0$, and unsharp measurements are in between.
The measurement operators $\Mo_\nu$ describing an unsharp measurement read as:
\begin{equation}\label{eq:Kraus}
\Mo_\pm=\sqrt{\Eo_\pm}=\frac{\cos(\eta)\pm\sin(\eta)\sio_z}{\sqrtwo}
\end{equation}

For a photon initially in the state \eqref{eq:gps:HV} the outcomes  $\nu=\pm 1$ of the measurement $\Mo_\pm$ update the normalized polarization state to
\begin{equation}\label{eq:nps:update}
\ket{\psi'_{\pm}}=\frac{1}{\sqrt{p_\pm}}\ket{\psi_\pm}=\frac{1}{\sqrt{p_\pm}}\Mo_\pm\ket{\psi},
\end{equation}
where $\ket{\psi_\pm}$ is the unnormalized state after the measurement and
\begin{equation}\label{eq:nps:probs}
p_\pm=\langle\psi_\pm\vert\psi_\pm\rangle=\bra{\psi}\Eo_\pm\ket{\psi}
\end{equation}
are the probabilities of the given outcomes.

The outcomes $\nu=\pm1$ are not yet the measured values $\si_z$ of the observable $\sio_z$ since
\begin{equation}\label{}
\Mb\nu=p_+-p_- = \sin(2\eta)\bra{\psi}\sio_z\ket{\psi},
\end{equation}
where $\Mb$ stands for the statistical mean. 
The correct definition of the measured value $\si_z$ is calibrated as 
\begin{equation}\label{eqn:calibr}
\si_z=\frac{1}{\sin(2\eta)}\nu,
\end{equation}
to satisfy
\begin{equation}\label{eqn:13}
\Mb\si_z=\bra{\psi}\sio_z\ket{\psi}.
\end{equation}

Unsharp measurement in any other basis is straightforward: replace $\sio_z$
by $\sio_{\nv}=n_x\sio_x+n_y\sio_y+n_z\sio_z$ along the direction of the Bloch vector $\nv=[n_x,n_y,n_z]$ in question.
The observable $\sio_{\nv}$ can be derived from $\sio_z$ as
\begin{equation}\label{eq:sion}
    \sio_{\nv}=\Uo(\vec{n})\sio_z\Uo^\dagger(\vec{n})
\end{equation}
by the unitary rotation
\begin{equation}\label{eq:Uonv}
\Uo(\nv)=\cos(\theta/2)-i\sin(\theta/2)\frac{n_x\sio_y-n_y\sio_x}{\sqrt{1-n_z^2}},
\end{equation}
where $\cos(\theta)=n_z$.

\section{Sequential unsharp measurements of photon polarization}\label{sec:SUMOPP}
\label{SUM}
Let us consider the sequential unsharp measurements of the observables $\sio_k=\sio_{\nv_k}$ in a sequence $k=1,\dots,N$. 
Note that the Bloch vectors $\nv_k$ determining the observables $\sio_k$ can be generally different for each measurement.

Applying the formalism developed in the previous section, these sequential measurements can be described by the following effects $\Eo_{\nu_1\dots\nu_N}(\eta_1,\dots,\eta_N)$ and measurement operators $\Mo_{\nu_1\dots\nu_N}(\eta_1,\dots,\eta_N)$: 
\begin{multline}\label{Effectmulti}
\Eo_{\nu_1\dots\nu_N}(\eta_1,\dots,\eta_N)=\\
=\Mo^\dagger_{\nu_1\dots\nu_N}(\eta_1,\dots,\eta_N)\Mo_{\nu_1\dots\nu_N}(\eta_1,\dots,\eta_N),
\end{multline}                                                
where
\begin{equation}\label{Measopmulti}
\Mo_{\nu_1\dots\nu_N}(\eta_1,\dots,\eta_N)=\Mo^{(N)}_{\nu_N}(\eta_N)\cdots\Mo^{(1)}_{\nu_1}(\eta_1)
\end{equation}
and
\begin{equation}\label{Mk}
\Mo^{(k)}_{\nu_k}(\eta_k)=\frac{\cos(\eta_k)+\nu_k\sin(\eta_k)\sio_k}{\sqrtwo}.
\end{equation}
The quantities $\nu_k=\pm1$ for $k=1,\dots,N$ are the possible outcomes of the unsharp measurements.

The effects $\Eo_{\nu_1\dots\nu_N}(\eta_1,\dots,\eta_N)$ satisfy the following closure relation:
\begin{equation}\label{}
\sum_{\{\nu\}}\Eo_{\nu_1\dots\nu_N}(\eta_1,\dots,\eta_N)=1,
\end{equation}
with the notation $\sum_{\{\nu\}}\equiv\sum_{\nu_N=\pm1}\dots\sum_{\nu_1=\pm1}$.
For simplicity, in the following we use the notation $\Mo_{\nu_k}^{(k)}$ instead of $\Mo_{\nu_k}^{(k)}(\eta_k)$ for any values of $k=1,\dots,N$. We also omit the arguments of $\Mo_{\nu_1\dots\nu_N}(\eta_1,\dots,\eta_N)$ and $\Eo_{\nu_1\dots\nu_N}(\eta_1,\dots,\eta_N)$.

For a photon initially in the state \eqref{eq:gps:HV}, the outcomes  $\nu_1,\dots,\nu_N$ of the sequential unsharp measurement $\Mo_{\nu_1\dots\nu_N}$ update the normalized polarization states to \begin{equation}\label{eq:22}
\ket{\psi'_{\nu_1\dots\nu_N}}=\frac{1}{\sqrt{p_{\nu_1\dots\nu_N}}}\ket{\psi_{\nu_1\dots\nu_N}}=
\frac{1}{\sqrt{p_{\nu_1\dots\nu_N}}}\Mo_{\nu_1\dots\nu_N}\ket{\psi},
\end{equation}
where $\ket{\psi_{\nu_1\dots\nu_N}}$ is the unnormalized state after $N$ sequential unsharp measurements and
\begin{equation}\label{eq:21}
p_{\nu_1\dots\nu_N}=\langle\psi_{\nu_1\dots\nu_N}\vert\psi_{\nu_1\dots\nu_N}\rangle=\bra{\psi}\Eo_{\nu_1\dots\nu_N}\ket{\psi}
\end{equation}
are the joint probabilities of the given outcomes.

Note that the sequence of the outcomes $\nu_1,\dots,\nu_N$ is not yet the collection of measured values $\si_1,\dots,\si_N$ of the observables $\sio_1,\dots,\sio_N$.
The correct measured values are the calibrated ones 
\begin{equation}\label{eq:sigmak}
\si_{k}=\frac{1}{\sin(2\eta_k)}\nu_k,
\end{equation}
as it was in the case of the single measurement \eqref{eqn:calibr}.

Knowing the probabilities $p_{\nu_1\dots\nu_N}$ of particular measurement results defined in Eq.~\eqref{eq:21} and using Eq.~\eqref{eq:sigmak}, one can easily calculate any $m$th order correlation function:
\begin{multline}\label{corrNvar}
\Mb\si_{k_1}\cdots\si_{k_m}=\\
=\sum_{\{\nu\}}p_{\nu_1\dots\nu_N}\frac{\nu_{k_1}}{\sin(2\eta_{k_1})}\cdots\frac{\nu_{k_m}}{\sin(2\eta_{k_m})},
\end{multline}
where $\{k_1,\dots,k_m\}\subseteq\{1,\dots,N\}$.
Let us discuss first the full $N$th order correlation function.
By substituting Eqs.~\eqref{Effectmulti} and \eqref{Measopmulti} into \eqref{corrNvar} for $\{k_1,\dots,k_m\}=\{1,\dots,N\}$ we obtain
\begin{multline}
\Mb\si_1\si_2\cdots\si_N=\\
=\sum_{\{\nu\}}\bra{\psi}\Mo_{\nu_1}^{(1)}\Mo_{\nu_2}^{(2)}\cdots
\Mo_{\nu_N}^{(N)}\Mo_{\nu_N}^{(N)}\cdots \Mo_{\nu_1}^{(1)}\ket{\psi}
\times\\
\times\frac{\nu_1}{\sin(2\eta_1)}\frac{\nu_2}{\sin(2\eta_2)}\cdots\frac{\nu_N}{\sin(2\eta_N)}
\end{multline} 

First, we evaluate the sum over $\nu_1$ of the factors containing $\nu_1$.
If $[\dots]$ stands for the operator product depending on $\nu_2,\dots,\nu_N$,
we have
\begin{eqnarray}
\sum_{\nu_1=\pm1}\Mo_{\nu_1}^{(1)}[\dots]\Mo_{\nu_1}^{(1)}\frac{\nu_1}{\sin(2\eta_1)}
&=&\frac{1}{2}\left(\sio_1[\dots]~+~[\dots]\sio_1\right)\nonumber\\
&\equiv&\frac{1}{2}\{\sio_1,[\dots]\},
\end{eqnarray}
where we used $\Mo_{\nu_1}^{(1)}=(\cos(\eta_1)+\nu_1\sin(\eta_1)\sio_1)/\sqrtwo$ 
according to Eq.~\eqref{Mk}. The symbols $\{~,~\}$ denote the anticommutator. 
Observe that the sharpness parameter $\eta_1$ is canceled.
Repeating the above summation over 
$\nu_2,\dots,\nu_N$ step by step, we arrive at the final result:
\begin{equation}\label{corrNanticomm}
\Mb\si_1\si_2\cdots\si_N
=2^{-N}\bra{\psi}\{\sio_1,\{\sio_2,\dots\{\sio_{N-1},\sio_N\}\}\dots\}\}\ket{\psi}
\end{equation}
The result, independent of the sharpness parameters $\eta_k$, takes the ultimate simple form 
in terms of the Bloch-vectors of the measured polarizations:
\begin{equation}\label{eqn:27}
\Mb\si_1\si_2\cdots\si_N
=\left\{
\begin{array}{lc}
(\nv_1\nv_2)(\nv_3\nv_4)\cdots(\nv_{N-1}\nv_N)
&\mbox{even }N\\
\bra{\psi}\sio_1\ket{\psi}(\nv_2\nv_3)\cdots(\nv_{N-1}\nv_N)
&\mbox{odd }N
\end{array}\right.
\end{equation}
The above method of derivation can be readily applied to any $m$th order correlation functions $\Mb\si_{k_1}\cdots\si_{k_m}$ defined in Eq.~\eqref{corrNvar}.
In this case, the sharpness parameters $\eta_k$ of the sequential measurement will also be cancelled.
Accordingly, measured correlations are independent of whether we measure the polarizations in projective, unsharp, or even weak measurements.
At the same time, one should know that the smaller the sharpnesses $\eta_k$ the larger the needed statistics are to  get the left-hand side of \eqref{corrNanticomm} reliably.
Similar result and the expressions (\ref{corrNanticomm},\ref{eqn:27}) were derived in Ref.~\cite{Dio16}, whereas the model of unsharp measurements was different from our binary-outcome measurements.

\begin{figure}[t!]
\includegraphics[width=0.6\columnwidth]{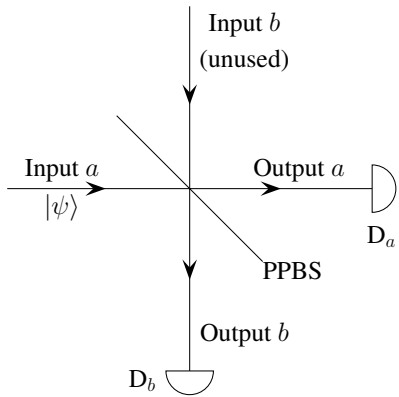}
    \caption{Schematic diagram of a setup realizing unsharp measurement of the photon polarization. A single photon in the polarization state $\ket{\psi}$ arrives at the PPBS in input mode $a$. Input mode $b$ is unused. The photon after the PPBS is detected either by the detector D$_a$ or by the detector D$_b$.}
    \label{fig:BeamSplitter}
\end{figure}
\section{Realization of sequential unsharp measurement}
\label{SUMreal}
A convenient way of realizing an unsharp measurement of the photon polarization is encoding some part of the polarization information of the photon into a path degree of freedom and detecting the photon in the given path. The proposed idea can be realized most straightforwardly by a PPBS. This is a two-mode optical device that
preserves the polarizations $H,V$ of the incident modes and transforms their paths $a,b$
into the outgoing modes differently for the horizontal and vertical polarization components for the input beams.
For a single photon, the four basis states transform like this:
\begin{equation}
              \begin{pmatrix}\ket{a}\\ \ket{b}\end{pmatrix}_{\text{out}}\!\!\!\ket{H/V}
=\hat{U}_{H/V}\begin{pmatrix}\ket{a}\\ \ket{b}\end{pmatrix}_{\text{in}}\!\!\!\ket{H/V},
\end{equation}
where the $2\times2$ unitary matrix is
\begin{equation}
    \hat{U}_{H/V}=\begin{pmatrix}
        \sqrt{T_{H/V}}e^{i\phi_{H/V}}&\sqrt{R_{H/V}}e^{i\psi_{H/V}}\\
        -\sqrt{R_{H/V}}e^{-i\psi_{H/V}}&\sqrt{T_{H/V}}e^{-i\phi_{H/V}}
    \end{pmatrix}. \label{eq:BSU}
\end{equation}

In Eq.~\eqref{eq:BSU} $T_{H/V}$ and $R_{H/V}$ are the transmission and reflection coefficients, respectively, satisfying the equation $T_{H/V}+R_{H/V}=1$.
The values of these coefficients can be chosen independently for the horizontal and vertical components.
Throughout our calculations the phases $\phi_{H/V}$ and $\psi_{H/V}$ are chosen to be 0.
We note that the transformations in Eq.~\eqref{eq:BSU} can describe PPBSs realized as bulk optical devices using multilayer dielectric coating and also those built by using various interferometric setups \cite{Floetal18}.

Let us consider first a scheme presented in Fig.~\ref{fig:BeamSplitter} containing a PPBS and two detectors. A single photon in a general polarization state $\ket{\psi}$ defined in Eq.~\eqref{eq:gps:HV} enters the setup in mode $a$, while port $b$ is unused. Assume a PPBS for which the transmission coefficient $T_H$ of the horizontal polarization
coincides with the reflection coefficient $R_V$ of the vertical polarization
and vice versa, that is, $T_H=R_V=\cos^2(\chi)$ and $R_H=T_V=\sin^2(\chi)$.
Introducing the basis $\{\ket{a},\ket{b}\}$ for the possible paths of the photon and using Eq.~\eqref{eq:BSU} the PPBS realizes the following transformation:
\begin{equation}
\label{eq:PPBStransform}
\begin{aligned}
\ket{a}\ket{H}&\Longrightarrow&\bigl(\cos(\chi)\ket{a}+\sin(\chi)\ket{b}\bigr)\ket{H},\\
\ket{a}\ket{V}&\Longrightarrow&\bigl(\sin(\chi)\ket{a}+\cos(\chi)\ket{b}\bigr)\ket{V}.
\end{aligned}
\end{equation}

Hence, by applying Eq.~\eqref{eq:Pauli3}, 
the transformation for a general polarization state $\ket{\psi}$ of Eq.~\eqref{eq:gps:HV} arriving at the PPBS on the path $\ket{a}$ reads
\begin{widetext}
\begin{eqnarray}\label{eq:trans:gps}
\ket{a}\ket{\psi}\Longrightarrow&&
\left\{\bigl(\cos(\chi)\ket{a}+\sin(\chi)\ket{b}\bigr)\frac{1+\sio_z}{2}
         +\bigl(\sin(\chi)\ket{a}+\cos(\chi)\ket{b}\bigr)\frac{1-\sio_z}{2}\right\}\ket{\psi}
         \nonumber\\
&=&\left\{\ket{a}\Mo_+ + \ket{b}\Mo_-\right\}\ket{\psi},
\end{eqnarray}
\end{widetext}
where $\Mo_\pm$ are defined in Eq.~\eqref{eq:Kraus}.
From this expression one can conclude that detecting the single-photon in mode $a$ or $b$ corresponding to the projective measurements $\ket{a}\bra{a}$ or $\ket{b}\bra{b}$, respectively, an unsharp measurement of the polarization of the photon can be realized by the scheme of Fig.~\ref{fig:BeamSplitter}.
Indeed, the collapse on $\ket{a}$ corresponds to the outcome $\nu=+1$ of the polarization measurement while the collapse on $\ket{b}$ means $\nu=-1$.
Introducing the pertinent notations $\ket{a}=\ket{+}$ and $\ket{b}=\ket{-}$, the full output state $\ket{\Psi}$ of the photon after the PPBS in Eq.~\eqref{eq:trans:gps} can be written as
\begin{eqnarray}\label{eq:Psi}
\ket{\Psi}&=&\left(\sum_{\nu=\pm1}\ket{\nu}\Mo_{\nu}\right)\ket{\psi}.
\end{eqnarray}
The resulting normalized polarization states $\ket{\psi'_{\pm}}$ of the photon corresponding to the outcomes $\nu=\pm1$ are really the ones defined in Eqs.~\eqref{eq:nps:update} and \eqref{eq:nps:probs}.

We note that the unsharp measurement of the polarization of the photon in a basis determined by the Bloch vector $\vec{n}=[n_x,n_y,n_z]$ corresponding to the measurement of the observable $\sio_{\vec{n}}$ can be realized in practice by placing the suitable wave plates before and after the PPBS in Fig.~\ref{fig:BeamSplitter} realizing the corresponding rotations of the Bloch vector described in Eqs.~\eqref{eq:sion} and \eqref{eq:Uonv}.

\begin{figure}[t]
\includegraphics[width=\columnwidth]{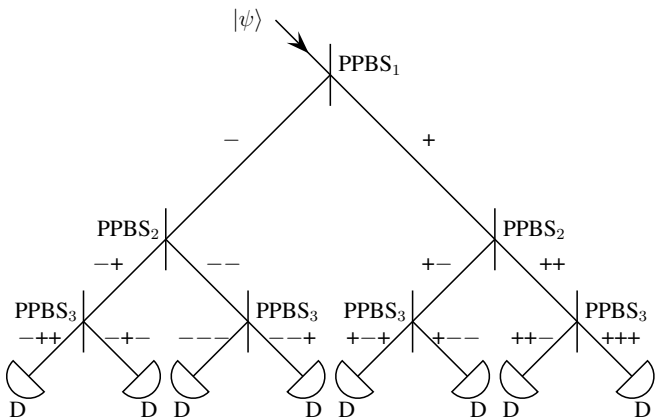}
\caption{Binary tree of PPBSs realizing sequential unsharp measurements of photon polarization for $N=3$ measurements.
The $k$th level of the binary tree corresponds to the $k$th unsharp measurement $\Mo_{\nu_k}^{(k)}$. PPBSs on level $k$ denoted by PPBS$_k$ are identical but they can be different on different levels.
Pre- and post-PPBS wave-plates necessary for measuring the observable $\sio_k$ are merged into PPBS symbols.
Possible paths of the photon are labeled by strings of the outcomes $\nu_1\dots\nu_k$ corresponding to the particular sequences of the measurement.
Transmission/reflection at a PPBS yields $+/-$ in the string, respectively. 
The input photon polarization state $\ket{\psi}$  of the setup is defined in Eq.~\eqref{eq:gps:HV}. Single-photon detectors are denoted by D.}
\label{fig:3levelbtree}
\end{figure}

Inspired by these considerations concerning the setup of Fig.~\ref{fig:BeamSplitter} containing a single PPBS, we propose using an $N$-level binary tree of PPBSs for the realization of sequential unsharp measurements of photon polarization, that is, the observables $\sio_k$ in a sequence $k=1,\dots,N$.
Recall that these sequential measurements can be described by the measurement operators $\Mo_{\nu_1\dots\nu_N}(\eta_1,\dots,\eta_N)$ defined in Eqs.~\eqref{Measopmulti} and \eqref{Mk}. An example of such a setup is presented in Fig.~\ref{fig:3levelbtree} for $N=3$.
The $k$th level of the binary tree corresponds to the $k$th unsharp measurement described by the measurement operator $\Mo_{\nu_k}^{(k)}$. PPBSs on level $k$ denoted by PPBS$_k$ are identical but they can be different on different levels.
Pre- and post-PPBS wave-plates necessary for measuring the observable $\sio_k$ are merged into PPBS symbols.
Possible paths of the photon are labeled by strings of the measurement outcomes $\nu_1\dots\nu_k$.
A $+$ sign in these strings means transmission of the photon at a PPBS of the given level while the $-$ sign means reflection.

Applying Eq.~\eqref{eq:Psi} repeatedly to all the levels of the binary tree setup, the full output state $\ket{\Psi^{(k)}}$ at the $k$th level of an $N$-level binary tree setup realizing a sequential unsharp measurement of the photon polarization can be written as
\begin{multline}\label{}
\ket{\Psi^{(k)}}=\sum_{\nu_1,\dots,\nu_k=\pm1}\left(\ket{\nu_1\dots\nu_k}\Mo^{(k)}_{\nu_k}\cdots\Mo^{(1)}_{\nu_1}\right)\ket{\psi}=\\
=\sum_{\nu_1,\dots,\nu_k=\pm1}\ket{\nu_1\dots\nu_k}\ket{\psi_{\nu_1\dots\nu_k}},
\end{multline}
where $k=1,\dots,N$. The state $\ket{\psi}$ is the initial polarization state \eqref{eq:gps:HV} and
\begin{equation}\label{eq:psin1nk}
\ket{\psi_{\nu_1\dots\nu_k}}=\Mo^{(k)}_{\nu_k}\cdots\Mo^{(1)}_{\nu_1}\ket{\psi}= \Mo_{\nu_1\dots\nu_k}\ket{\psi}
\end{equation}
is the unnormalized polarization state of the photon travelling on the path labeled by the string $\nu_1\dots\nu_k$.
If the photon is detected after the $N$th level of PPBSs then the joint probability of counts can be calculated as
\begin{equation} \label{eq:pn1nk}
p_{\nu_1\dots\nu_N}=\langle\psi_{\nu_1\dots\nu_N}\ket{\psi_{\nu_1\dots\nu_N}}.
\end{equation}
The unsharpness parameters $\chi_k$ of the measurements $\Mo_{\nu_k}^{(k)}$ realized by the $k$th level of the binary tree setup are determined by the horizontal and vertical transmission and reflection coefficients characterizing the PPBSs of the given level as
\begin{eqnarray}
T_H^{(k)}=R_V^{(k)}=\cos^2(\chi_k),\label{eq:THk}\\
R_H^{(k)}=T_V^{(k)}=\sin^2(\chi_k).\label{eq:RHk}
\end{eqnarray}

Comparing Eqs.~\eqref{eq:psin1nk} and \eqref{eq:pn1nk} with Eqs.~\eqref{eq:22} and \eqref{eq:21} one can conclude that the proposed scheme indeed realizes a sequential unsharp measurement described in Sec.~\ref{sec:SUMOPP}.
We note that the probabilities \eqref{eq:pn1nk} can also be considered as classical, that is, they describe the detected intensities when the initial state is a classical
polarized light instead of a single photon. The setup becomes essentially quantum only for single-photon input with postselection. We shall discuss this property in detail later.

Using Eqs.~\eqref{Mk}, \eqref{eq:PPBStransform}, \eqref{eq:THk}, and \eqref{eq:RHk}, the unnormalized polarization state $\ket{\psi_{\nu_1\dots\nu_N}}$ for the case of $N$ unsharp measurements of the observable $\sio_z$ can be expressed as
\begin{eqnarray}
\ket{\psi_{\nu_1\dots\nu_N}}=&\alpha&\prod_{k=1}^N\cos^{t_k}(\chi_k)\sin^{1-t_k}(\chi_k)\ket{H}\nonumber\\
      +&\beta&\prod_{k=1}^N\cos^{1-t_k}(\chi_k)\sin^{t_k}(\chi_k)\ket{V}
\end{eqnarray}
and the probabilities corresponding to a particular measurement result determined by the string $\nu_1\dots\nu_N$ read
\begin{eqnarray}\label{eq:38}
p_{\nu_1\dots\nu_N}=&\vert\alpha\vert^2&\prod_{k=1}^N\cos^{2t_k}(\chi_k)\sin^{2(1-t_k)}(\chi_k)\nonumber\\
       +&\vert\beta\vert^2&\prod_{k=1}^N\cos^{2(1-t_k)}(\chi_k)\sin^{2t_k}(\chi_k)
\end{eqnarray}
where
\begin{equation}
    t_k=\begin{cases}
    1 \text{ if }\nu_k=+1\\
    0 \text{ if }\nu_k=-1
    \end{cases},
\end{equation}
that is, $t_k=1$ accounts for the photon's transmission ($t_k=0$ does for reflection) in a PPBS in the $k$th level of the setup along the path determined by the string $\nu_1\dots\nu_N$.

Using Eqs.~\eqref{corrNvar} and \eqref{eq:38} one can easily derive that the $m$th order correlation functions are
\begin{equation}
\Mb\si_{k_1}\cdots\si_{k_m}=\begin{cases}
\vert\alpha\vert^2-\vert\beta\vert^2, &\text{ for odd } m,\\
1, &\text{ for even } m.\\
\end{cases}
\end{equation}
In accordance with the considerations after Eq.~\eqref{corrNvar} in Sec.~\ref{sec:SUMOPP}, the unsharpness parameters do not appear in these formulas. As we will show in the next section these simple relationships become non-trivial if postselection is introduced into the scheme.

\section{Unsharp measurements with post- and reselection}
\label{UMpsrs}
Our general theoretical description of sequential unsharp measurement  of photon polarization presented in Sec.~\ref{sec:SUMOPP} can be applied for arbitrary sharpness parameters.
Hence it can also be applied for WMs that correspond to the limit $\eta_k\to 0$ of the sharpness parameters.
In this limit backreaction of the unsharp measurement on the polarization vanishes, the measurement becomes noninvasive, the measured polarization state remains undisturbed.
In the experimental scheme of Fig.~\ref{fig:3levelbtree} realizing a sequential unsharp measurement the WM limit
assumes the use of PPBSs corresponding to very small values of $\eta_k$.

In the theory of WM the concepts of postselection and reselection have crucial role. 
They predict anomalous measured values \cite{AAV88} and correlation functions \cite{Dio16}. 
In this section we analyze the consequences of using post\-selection and reselection in our theory.

\subsection{Anomalous mean value at postselection}

Consider the measurement of a given observable $\sio$ corresponding to a direction $\vec{n}$ of the polarization for an initial polarization state $\ket{i}=\ket{\psi}$.
Repeating the measurement for the same initial state many times one can obtain the statistics of the measured values $\si$.
Then the average of the measured values $\Mb\si$ satisfies
\begin{equation}
\Mb\si=\bra{i}\sio\ket{i}.\label{eq:41}
\end{equation}
This expression is valid for any unsharp measurement with arbitrary sharpness parameter (cf.\ Eqs.~\eqref{eqn:13} and \eqref{eqn:27}), and obviously for a weak measurement, too.

However, the original concepts of WM implemented postselection \cite{AAV88}.
This idea assumes the addition of a projective measurement after the WM and discarding those outcomes for which the projective measurement does not yield a certain final state $\ket{f}$.
In this case the theoretical prediction for the average measured values becomes radically different:
\begin{equation}\label{eq:42}
\Mb\si\vert_\text{psel}=\mathrm{Re}\frac{\bra{f}\sio\ket{i}}{\langle f\ket{i}}\equiv\mathrm{Re}\si^\mathrm{WV},
\end{equation}
where the ratio is called the complex weak value and denoted by $\si^\mathrm{WV}$.
Since the denominator can be arbitrary small, the measured value $\Mb\si\vert_\mathrm{psel}$  may fall far away from the spectrum $\pm1$ of $\sio$ \cite{AAV88}.

Inspired by this result, let us consider a general unsharp measurement with postselection.
In our theory this corresponds to two consecutive measurements.
The first measurement is the unsharp one on $\sio_1=\sio$, of sharpness $\eta_1=\eta$, described by measurement operators $\Mo^{(1)}_{\nu_1}=2^{-1/2}(\cos(\eta)+\nu_1\sin(\eta)\sio)$.
The second measurement on $\sio_2=\ket{f}\bra{f}-\ket{f_\perp}\bra{f_\perp}$ is projective (sharp, $\eta_2=\pi/4$),
described by measurement operators $\Mo^{(2)}_{\nu_2}(\pi/4)$. At successful postselection we only need one of them: $\Mo^{(2)}_+=\ket{f}\bra{f}$.
Using Eq.~\eqref{eq:psin1nk}, the unnormalized polarization state $\ket{\psi_{\nu_1+}}$ after the two measurements reads as
\begin{equation}\label{eqn:44}
\ket{\psi_{\nu_1+}}=2^{-1/2}\ket{f}\bra{f}(\cos(\eta)+\nu_1\sin(\eta)\sio)\ket{i}.
\end{equation}
Applying Eq.~\eqref{eq:pn1nk}, the probability of the given outcomes of the measurements can be calculated as
\begin{eqnarray}\label{eqn:45}
p_{\nu_1+}&=&\langle{\psi_{\nu_1+}}\ket{\psi_{\nu_1+}}=\frac{1}{2}\vert\langle{f}\ket{i}\vert^2\times\\
&\times&\left(\cos^2(\eta)+\vert\si^{\mathrm{WV}}\vert^2\sin^2(\eta)+\nu_1\si^{\mathrm{WV}}\sin(2\eta)\right).
\nonumber
\end{eqnarray}
The average of the measured values $\si=\nu_1/\sin(2\eta)$ of the polarization is
\begin{equation}\label{eqn:46}
\Mb\si\vert_\mathrm{psel}=\frac{1}{\sin(2\eta)}\frac{\sum_{\nu_1=\pm}\nu_1p_{\nu_1+}}{\sum_{\nu_1=\pm}p_{\nu_1+}}
\end{equation}
yielding
\begin{equation}\label{eqn:47}
\Mb\si\vert_\mathrm{psel}=\frac{\mathrm{Re}\si^\mathrm{WV}}{\cos^2(\eta)+\vert\si^\mathrm{WV}\vert^2\sin^2(\eta)}.
\end{equation} 
This expression is valid for any value of the sharpness parameter $\eta$.
In WM limit $\eta\rightarrow0$, it converges to the theoretical value $\mathrm{Re}\si^{\mathrm{WV}}$ in Eq.~\eqref{eq:42} indeed.

The anomalous mean value in Eq.~\eqref{eqn:47} can be measured by using our general experimental scheme in the following way.
To realize an unsharp measurement with postselection, one needs a two-level version of the binary-tree system presented in Fig.~\ref{fig:3levelbtree}.
Using Eqs.~\eqref{eq:THk} and \eqref{eq:RHk}, the transmission $R_{H/V}$ (and reflection $T_{H/V}$) coefficients of the PPBS on the first level should be chosen to yield small sharpness parameter $\eta\ll\pi/4$, while the two PPBSs on the second level should realize a projective measurement, they must have $\eta=\pi/4$. Accordingly, the latter two PPBSs are eventually PBSs.
The measurement of $\sio$ corresponding to a particular direction $\vec{n}$ may require waveplates before and after the PPBS. Similarly, the two PBSs will be sandwitched by  waveplates ensuring that the projective measurement be in the basis 
$\ket{f},\ket{f_\perp}$.
We note that it is enough using only those two detectors in this scheme that are placed in the paths corresponding to the chosen postselected polarization state $\ket{f}$.

\subsection{Anomalous correlation at reselection}
Reselection \cite{Dio16} is the special case of postselection when the final state $\ket{f}$ is equal to the initial state $\ket{i}$, that is, $\ket{f}=\ket{i}$.
Comparing Eqs.~\eqref{eq:41} and \eqref{eq:42} shows that in this case the average of the measured value $\si$ is the same
no matter without or with reselection. This is what we expect since the ideal WM does not alter the initial state $\ket{i}$.
Indeed, reselection does not matter for a single WM.
For multiple WMs, however, an unexpected new anomaly pops up even for the simplest correlation.
Suppose one performs the WM of the same polarization $\sio$ twice, yielding the measured values $\si_1,\si_2$, respectively.
In Ref.~\cite{Dio16} it was shown that in the case of reselection, the second order correlation function $\Mb\si_1\si_2\vert_\text{resel}$ for this sequential WM of photon polarization is
\begin{equation}\label{eqn:48}
\Mb\si_1\si_2\vert_\mathrm{resel}=\frac{1}{2}\bigl(1+\langle\sio\rangle^2\bigr).
\end{equation}
Without any postselection this correlation function would take the trivial value
\begin{equation}
\Mb\si_1\si_2=\bra{i}\sio^2\ket{i}=1.
\end{equation}
Let us first consider the second-order correlation function $\Mb\si_1\si_2\vert_{\text{resel}}$ for the case when two consecutive unsharp measurements are used instead of two WMs, which latter we shall consider afterwards. 
In our formalism, $\sio_1=\sio_2=\sio$, $\sio_3=\ket{i}\bra{i}-\ket{i_\perp}\bra{i_\perp}$ while $\eta_1=\eta_2=\eta$,
$\eta_3=\pi/4$.
Similarly to the derivation of Eqs.~\eqref{eqn:44} and \eqref{eqn:45}, after the two identical unsharp measurements of $\sio$
and the projective measurement corresponding to reselection of $\ket{i}$, the unnormalized polarization state reads
\begin{multline}
\ket{\psi_{\nu_1\nu_2+}}=\\
=\frac{1}{2}\ket{i}\bra{i}\left(\cos(\eta)+\nu_2\sin(\eta)\sio\right)\left(\cos(\eta)+\nu_1\sin(\eta)\sio\right)\ket{i}.
\end{multline}
The corresponding probability of the given outcomes of the measurements can be calculated as
\begin{multline}
p_{\nu_1\nu_2+}=\langle{\psi_{\nu_1\nu_2+}}\ket{\psi_{\nu_1\nu_2+}}=
\frac{1}{4}+\frac{\nu_1+\nu_2}{4}\sin(2\eta)\langle\sio\rangle\\
+\frac{\sin^2(2\eta)}{8}\left(\langle\sio\rangle^2-1+\nu_1\nu_2(\langle\sio\rangle^2+1)\right).
\end{multline}
The second-order correlation function of the measured values $\si_1=\nu_1/\sin(2\eta)$ and $\si_2=\nu_2/\sin(2\eta)$ of the polarization, that is, the average of their product is
\begin{equation}
\Mb\si_1\si_2\vert_\mathrm{resel}
=\frac{1}{\sin^2(2\eta)}\frac{\sum_{\nu_1,\nu_2=\pm}\nu_1\nu_2p_{\nu_1\nu_2+}}{\sum_{\nu_1,\nu_2=\pm}p_{\nu_1\nu_2+}}
\end{equation}
yielding
\begin{equation}\label{eqn:53}
 \Mb\si_1\si_2\vert_\mathrm{resel}
 =\frac{1+\langle\sio\rangle^2}{2-\sin^2(2\eta)(1-\langle\sio\rangle^2)}.
\end{equation}
This expression is the generalization of Eq.~\eqref{eqn:48} for the case of two consecutive unsharp measurements. 
In the WM limit $\eta\to0$ it converges to the expression In Eq.~\eqref{eqn:48}.

For measuring the anomalous correlation function in Eq.~\eqref{eqn:53}, one needs a three-level experimental scheme, 
like in Fig.~\ref{fig:3levelbtree}.  The first two levels are built by appropriate PPBSs realizing the consecutive unsharp measurements of the observable $\sio$ while the third level is built by four PBSs each realizing projective measurement in the polarization basis $\{\ket{i},\ket{i_{\perp}}\}$ corresponding to the initial state. Four single-photon detectors in the four reselected paths are sufficient.

A simple demonstration of this measurement in the WM limit can be the following.
We opt for the diagonal input (and the reselected) state as $\ket{i}=\ket{f}=(\ket{H}+\ket{V})\sqrt{2}$.
We measure $\sio=\sio_z=\ket{H}\bra{H}-\ket{V}\bra{V}$ twice at the same low sharpness $\eta\ll\pi/4$.
Finally, reselect on the diagonal state.  Since $\langle\sio\rangle=\bra{i}\sio_z\ket{i}=0$,  the measured 
correlation function in Eq.~\eqref{eqn:53} reduces to
\begin{equation}
 \Mb\si_1\si_2\vert_\mathrm{resel}
 =\frac{1}{2-\sin^2(2\eta)}
\end{equation}
which tends to the anomalous value $1/2$ for $\eta\to0$,
half of the naively expected value $1$.

\section{Conclusion}\label{Sec:conc}
We have proposed a general experimental scheme for realizing sequential unsharp measurements of photon polarization in which the sharpnesses and the bases of the particular photon polarization measurements can be chosen arbitrarily by using corresponding PPBSs and phase plates in the setup.
The scheme can also realize sequential WMs in the limit of low sharpnesses determined by the constituent PPBSs.
We have developed a general formalism describing this scheme in which the particular unsharp measurements are characterized by the appropriate measurement operators.
Our framework can be used to calculate the output polarization states after the sequential measurement and any correlation functions characterizing the measurement results, and also for analyzing the consequences of applying postselection and reselection in the measurement.
By using the proposed scheme, the anomalous mean value for an unsharp polarization measurement with postselection and the anomalous second-order correlation function for two consecutive measurements of photon polarization with reselection could be measured with ease.

\acknowledgments{This research was supported by the National Research, Development and Innovation Office, Hungary (Project Nos.\ TKP 2021-NVA-04, K124351,
and the ``Frontline'' Research Excellence Programme Grant No.\ KKP133827).}


%

\end{document}